\documentclass[12pt]{article}

\usepackage[margin=1in]{geometry}
\usepackage{graphicx}
\usepackage{amsmath}
\usepackage{authblk}
\usepackage{caption}
\usepackage{pdflscape}

\title{A scarab chart for visualizing mass patterns in light-hadron multiplets}
\author{R. M. Marinaro III}
\affil{School of Engineering and Computing, Christopher Newport University, Newport News, VA, USA}
\date{\today}

\begin{document}
\maketitle

\begin{abstract}
Particle charts are useful teaching tools, but the usual quark-model multiplet diagrams emphasize symmetry labels more than mass ordering. This paper presents a compact ``scarab chart'' for light hadrons that keeps the familiar isospin and hypercharge information while adding a mass-sensitive ordering. The chart is intended for modern physics or introductory particle physics teaching. Each tile represents a light meson or baryon, with columns labelled by isospin projection, rows ordered by an integer pion-anchored mass count, tile colour and faint pattern identifying hypercharge, and border colour and line style corresponding to the pseudoscalar nonet, vector-meson nonet, baryon octet, or baryon decuplet. The chart is proposed as a pedagogical visualization. The result gives students a single visual diagram in which quark content, multiplet, strangeness, and approximate mass scale can be compared.
\end{abstract}

\noindent\textbf{Keywords:} particle physics education; hadron multiplets; quark model; isospin; data visualization; particle diagrams

\section{Introduction}

The light-hadron spectrum is a useful entry point into particle physics because it connects observable particle properties with symmetry, quark content, and mass. Historically, the organization of mesons and baryons into multiplets was one of the central successes of the Eightfold Way and flavor \(SU(3)\) symmetry \cite{gellmann1961,neeman1961}. In these diagrams, particles are arranged according to quantum numbers such as isospin projection \(I_3\) and hypercharge \(Y\). The later quark model gave these multiplets a simple constituent interpretation where mesons are described by quark--antiquark valence structures, while baryons are described by three-quark valence structures \cite{gellmann1964,zweig1964}. These ideas remain standard topics in modern physics and introductory particle physics courses \cite{griffiths,perkins}.

\begin{figure}[h!]
\centering
\IfFileExists{scarab_chart_mixed_identified.pdf}{%
\includegraphics[angle=0,origin=c,width=0.72\textheight]{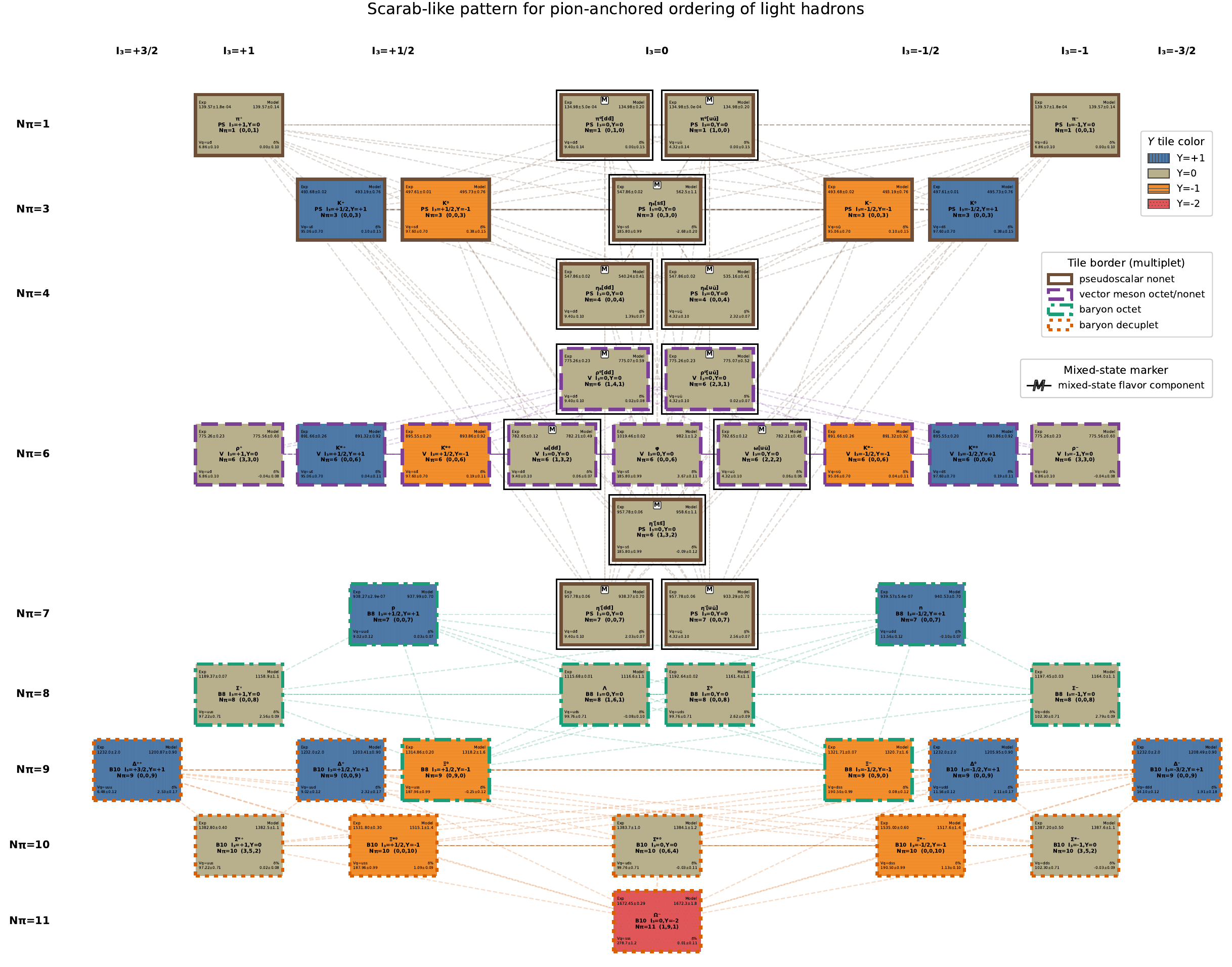}%
}{%
\IfFileExists{scarab_chart_mixed_identified.png}{%
\includegraphics[angle=0,origin=c,width=0.90\textheight]{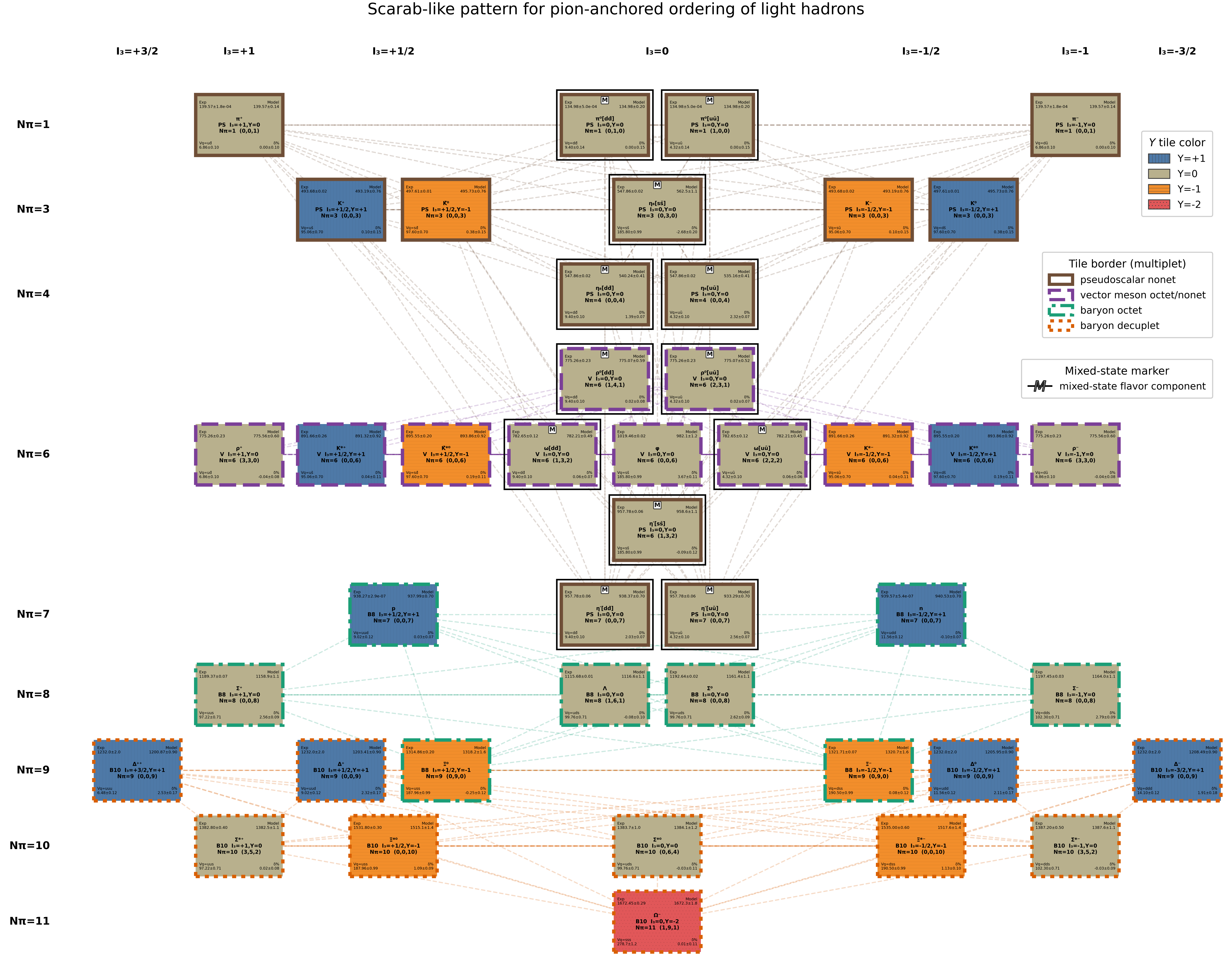}%
}{%
\fbox{\parbox{0.88\linewidth}{\centering Placeholder for \texttt{scarab\_chart.pdf}.\\ Rename the final chart file to \texttt{scarab\_chart.pdf} or update the filename in this figure.}}%
}%
}
\caption{A scarab chart for light-hadron multiplets. Columns give \(I_3\), rows give the integer pion count \(N_\pi\), and tile fill colour and pattern give hypercharge \(Y\). Border colour and line style give the multiplet class. A solid black outer border and the label ``M'' identify displayed flavor components of mixed neutral mesons. Dashed lines connect members of the same class.}
\label{fig:scarab}
\end{figure}

For students, however, the usual representations separate several pieces of information that are physically connected. A standard \(I_3\)-\(Y\) multiplet diagram shows charge, strangeness, and symmetry structure clearly, but mass values are usually supplied separately in a table. Conversely, a table of particle masses gives quantitative information, but it does not immediately show how the particles sit within nonets, octets, and decuplets. This separation can make the hadron spectrum appear as a list to memorize rather than as a structured pattern to interpret. The Particle Data Group tables provide the authoritative numerical data, but they are not designed as classroom visualizations of the spectrum \cite{pdg2026}. The scarab chart shown in Figure~\ref{fig:scarab} is designed to bridge several representations. It keeps the familiar horizontal organization by \(I_3\), uses tile colour and faint pattern to display hypercharge \(Y\), and uses border colour, border line style, and connecting lines to identify the pseudoscalar nonet, vector-meson nonet, baryon octet, and baryon decuplet. Its additional feature is a vertical coordinate based on a simple pion-anchored integer mass count, so that approximate mass ordering becomes part of the visual structure.

Research in physics education has shown that student reasoning and
problem-solving performance can depend strongly on the representation in
which physical information is presented. Coordinated verbal, pictorial,
graphical, and mathematical representations can support conceptual
understanding when their relationships and distinct purposes are made
explicit \cite{kohl2005,vanheuvelen2001,fredlund2012}. More generally,
multiple representations are most effective when they provide complementary
information and when learners are supported in relating one representation
to another \cite{ainsworth2006}. Particle physics education research
has likewise found that carefully designed typographic illustrations can
support students in distinguishing quarks, composite particle systems, and
the model-based character of descriptions of matter \cite{wiener2015}. The
scarab chart follows these principles by combining quantum-number position,
colour and pattern, multiplet-border coding, quark-content labels, and
measured mass while assigning each visual feature a distinct interpretive
role.

\section{Construction of the chart}

The scarab chart begins from the measured particle mass and separates it into a valence-quark part and a remaining mass content. For a hadron with assigned valence content \(q_i\),
\begin{equation}
M_{\rm NV}=M_{\rm exp}-\sum_i m_{q_i},
\end{equation}
where \(M_{\rm exp}\) is the experimental mass and \(m_{q_i}\) are light current-quark masses. The quantity \(M_{\rm NV}\) is not a dynamical contribution to the hadron mass. It is a bookkeeping quantity used to compare different light hadrons on a common mass-sensitive scale. The reference scale is fixed by pion masses. For the neutral and charged pion channels, the corresponding valence-quark masses are subtracted to define three pion-anchored mass units,
\begin{align}
U_{uu} &= M_{\pi^0}-2m_u,\\
U_{dd} &= M_{\pi^0}-2m_d,\\
U_{\pm} &= M_{\pi^\pm}-(m_u+m_d).
\end{align}
Each hadron is then assigned an integer triple $(N_{uu},N_{dd},N_{\pm})$, where \(N_{uu}\) and \(N_{dd}\) count neutral pion-like units and \(N_{\pm}\) counts charged pion-like units. For the chart shown here, the displayed integer triple is chosen as the nearest pion-anchored estimate to \(M_{\rm NV}\) within the light-hadron plotting set where the integer assignment is chosen by comparing the valence-subtracted mass content with
\begin{equation}
M_{\rm NV}^{\rm model}
=
N_{uu}U_{uu}+N_{dd}U_{dd}+N_{\pm}U_{\pm}.
\end{equation}
The displayed mass estimate is then
\begin{equation}
M_{\rm model}
=
\sum_i m_{q_i}
+
N_{uu}U_{uu}+N_{dd}U_{dd}+N_{\pm}U_{\pm}.
\end{equation}
The residual $\Delta M=M_{\rm exp}-M_{\rm model}$ is used only as a diagnostic of how closely the integer assignment reproduces the measured mass. It is not used as a coordinate in the chart. The vertical coordinate is the total integer pion count, $N_\pi=N_{uu}+N_{dd}+N_{\pm}$. The horizontal coordinate is the isospin projection \(I_3\). Thus the chart keeps the familiar multiplet direction associated with isospin, while the vertical direction introduces a simple mass-sensitive ordering. Hypercharge \(Y\) is shown by the tile fill colour, and the multiplet class is shown by the tile border colour and dashed connection lines. For mixed neutral mesons, separate component channels are shown when useful for teaching. This allows students to see how the same physical mass can be compared with different valence assignments, such as non-strange and strange quark--antiquark components. The construction is intentionally simple, so that the figure can be used in classroom discussion without requiring a full treatment of QCD mass generation. The numerical search range, tie-breaking rule, uncertainty propagation, and declared assignment constraint are documented in Appendix~\ref{sec:integer-assignment}.

\section{Classroom learning activity}
\label{sec:classroom-activity}

The activity is intended for a modern physics or introductory particle physics course after students have encountered quarks and antiquarks, mesons and baryons, electric charge, strangeness, isospin projection, hypercharge, and the idea of flavor multiplets. It can be completed in approximately 20--25 minutes by individuals or small groups. A simplified classroom chart is shown in Appendix~\ref{app:simplified-chart}. It retains the particle symbol, valence label, measured mass, \(I_3\) column, \(N_\pi\) row, hypercharge colour and pattern, and multiplet border colour and line style, while omitting the model mass, integer triple, residual, and propagated uncertainties that appear in the complete chart. The activity is organized around identifying, comparing, and translating
among these visuals rather than reading the chart passively, because
representational competence requires students to connect the physical
meanings carried by different representational forms
\cite{kohl2005,fredlund2012,ainsworth2006}.

The learning goals are for students to distinguish the different visuals used in the chart, connect quark content with \(I_3\) and hypercharge, compare mass ordering across related multiplets, and identify which parts of the construction are exact labels and which are approximate organizations. The instructor first reviews the legend and emphasizes that \(N_\pi\) is an integer mass-ordering coordinate, not a claim that the hadron contains that number of physical pions. The label ``M'' and the solid black outer border are also introduced as markers for alternative flavor-component representations of a mixed neutral meson. Students then work through the following sequence.
\begin{enumerate}
\item Locate \(\pi^+\), \(K^+\), the proton, and \(\Delta^{++}\). For each state, record its valence content, \(I_3\), hypercharge, multiplet border, and \(N_\pi\) row. Students should identify that the horizontal position is controlled by \(I_3\), whereas the vertical position is controlled by the pion-anchored mass count.

\item Trace one complete isospin family, such as the three pions or the four \(\Delta\) states. Students compare the valence labels as one \(u\) quark is successively replaced by a \(d\) quark and describe how this changes \(I_3\) while leaving the family on a common or nearby mass-ordering row.

\item Compare a non-strange state with one or more strange partners in the same broad class. Suitable examples are \(\pi\) and \(K\), the nucleons and \(\Xi\) baryons, or the \(\Delta\), \(\Sigma^*\), \(\Xi^*\), and \(\Omega^-\) sequence. Students identify the change in hypercharge color and determine whether the additional strange-quark content is accompanied by movement toward a larger \(N_\pi\).

\item Compare the pseudoscalar and vector mesons with similar valence content, for example \(\pi\) with \(\rho\) or \(K\) with \(K^*\). Students then compare the baryon octet and decuplet at similar flavor content. The goal is to recognize that multiplet membership and mass ordering are correlated but are not the same classification.

\item Locate the tiles carrying the ``M'' marker. Students should explain why several flavor-component tiles can carry the same measured particle mass and why these tiles do not represent several newly observed particles. They then identify which quantity changes between the component tiles, including the assumed valence subtraction and therefore the selected pion-anchored assignment.

\item Using the complete chart, select three states and compare \(M_{\rm exp}\), \(M_{\rm model}\), and \(\Delta M\). Students identify one state for which the integer estimate is close to the measured mass and one for which the disagreement is visibly larger. They are asked whether a nonzero residual invalidates the chart or instead measures the limitation of the simplified ordering rule.
\end{enumerate}

A short whole-class discussion can be organized around the distinction between exact and approximate information. The particle identity, measured mass, valence assignment used for bookkeeping, \(I_3\), hypercharge, and stated multiplet membership are explicit inputs or labels. The integer triple, \(N_\pi\), and model mass are outputs of the pion-anchored construction. The expected conclusion is that the chart does not replace the conventional \(I_3\)-\(Y\) diagrams or provide a QCD derivation of the spectrum. Its educational role is to place quantum numbers, flavor content, multiplet structure, and an approximate mass ordering in one representation that can be interrogated and criticized. The activity can be assessed with a brief exit question: ``Choose two visual features of the scarab chart that represent exact classification information and two that represent approximate mass-pattern information. Explain the distinction in one or two sentences.'' A satisfactory response should classify \(I_3\), hypercharge, particle identity, and multiplet labels as classification information, while identifying \(N_\pi\), the integer triple, model mass, and residual as parts of the approximate mass-sensitive scarab chart structure.

\section{Discussion}

The scarab chart is intended for use after students have been introduced to quarks, antiquarks, mesons, baryons, charge, strangeness, isospin, and the basic idea of hadron multiplets. A short classroom activity can begin by asking students to locate familiar states such as the pions, kaons, nucleons, and \(\Delta\) baryons. Students can then identify which particles are connected by the same border colour and relate those connections to the pseudoscalar nonet, vector-meson nonet, baryon octet, and baryon decuplet. In this way, the chart provides a compact visual summary of the same
classification ideas that are usually introduced through separate multiplet
diagrams. Its use of an integrated qualitative representation follows
established physics education approaches in which diagrams and other visual
forms are used to organize physical relationships before or alongside more
quantitative analysis \cite{vanheuvelen2001,kohl2005}.

The main pedagogical value of the chart is that it separates, but displays together, three different organizing ideas. The horizontal coordinate is a quantum-number coordinate, given by the isospin projection \(I_3\). The tile fill colour is a hypercharge cue, and therefore also gives students a visual indication of strangeness. The vertical coordinate is a mass-sensitive coordinate, given by the integer pion count \(N_\pi\). This allows students to see that symmetry membership and approximate mass ordering are related, but not identical. For example, students can compare the pseudoscalar and vector mesons, or the baryon octet and decuplet, and then discuss how increasing strange-quark content changes the hypercharge colour, the position of the tile, and the mass scale.

The chart should be presented as an organizing representation rather than as a first-principles mass formula. The integer pion count is a visualization device. It should not be interpreted as a derivation of hadron masses from QCD. Current-quark masses are convention dependent, and the physical masses of light hadrons arise from strong-interaction dynamics, chiral symmetry breaking, electromagnetic effects, spin-dependent interactions, and flavor mixing. The residuals between the measured masses and the integer pion-anchored estimates are therefore expected to be nonzero. In teaching, this limitation can be useful. The chart gives students an accessible pattern to examine, while also motivating the question of why real hadron masses only approximately follow simple ordering rules. Used in this way, the scarab chart complements the standard quark-model and \(I_3\)-\(Y\) multiplet diagrams. It does not replace those diagrams. Instead, it adds a mass-pattern layer that helps students move from memorizing particle lists toward interpreting the light-hadron spectrum as a structured but approximate physical pattern.

\section{Conclusion}

The scarab chart offers a compact classroom representation of the light-hadron spectrum. By placing states according to \(I_3\) and an integer pion-anchored mass count, while also showing hypercharge and multiplet membership through colour and connecting lines, the chart brings together information that is usually spread across separate diagrams and tables. Its purpose is not to derive hadron masses, but to make visible the approximate mass patterns that accompany the familiar pseudoscalar, vector-meson, baryon-octet, and baryon-decuplet structures. Used alongside the standard quark-model and \(I_3\)-\(Y\) diagrams, the scarab chart can help students interpret the light-hadron spectrum as a structured physical pattern rather than as a disconnected list of particles.



\clearpage
\appendix

\section{Integer-assignment procedure}
\label{sec:integer-assignment}

The numerical inputs are fixed before any non-pion hadron is assigned an integer triple. The light current-quark masses used in the calculation are the \(\overline{\mathrm{MS}}\) values at \(\mu=2\,\mathrm{GeV}\), and the pion and hadron masses are taken from the Particle Data Group values \cite{pdg2026}. The quark and pion inputs used for the valence subtraction and pion anchors are listed in Table~\ref{tab:scarab-inputs}. The resulting units are
\begin{equation}
U_{uu}=130.6568\,\mathrm{MeV},\qquad
U_{dd}=125.5768\,\mathrm{MeV},\qquad
U_{\pm}=132.71039\,\mathrm{MeV}.
\end{equation}
These three quantities are fixed by the pion and current-quark inputs and are not fitted to the remaining hadron spectrum. For each hadron or displayed flavor component, the code first calculates
\begin{equation}
M_{\rm NV}=M_{\rm exp}-\sum_i m_{q_i}.
\end{equation}
The integer assignment is then obtained by a search over non-negative integer triples,
\begin{equation}
(N_{uu},N_{dd},N_{\pm})\in\{0,1,\ldots,16\}^3,
\end{equation}
using
\begin{equation}
(N_{uu},N_{dd},N_{\pm})_{\rm best}
=
\mathop{\arg\min}_{0\leq N_{uu},N_{dd},N_{\pm}\leq16}
\left|
M_{\rm NV}-N_{uu}U_{uu}-N_{dd}U_{dd}-N_{\pm}U_{\pm}
\right|.
\label{eq:integer-search}
\end{equation}
All three integers are therefore non-negative, and no fractional or negative pion units are allowed. The upper bound of 16 is larger than the assignments required by the plotted light-hadron set and serves only to make the finite search explicit.

If two candidate triples have the same residual to numerical precision, the program applies a fixed lexicographic tie-breaking rule. It first selects the triple with the smaller total count \(N_{uu}+N_{dd}+N_{\pm}\), then the smaller \(N_{\pm}\), then the smaller \(N_{dd}\), and finally the smaller \(N_{uu}\). Equivalently, candidate triples are ranked by
\begin{equation}
\left(
|\Delta M|,
N_{uu}+N_{dd}+N_{\pm},
N_{\pm},
N_{dd},
N_{uu}
\right),
\end{equation}
where the tuple is minimized from left to right. The \(\pi^0\), \(\pi^+\), and \(\pi^-\) assignments are fixed to \((1,0,0)\), \((0,0,1)\), and \((0,0,1)\), respectively, so that the three basis labels retain their defining pion interpretation.

\begin{table}[h!]
\centering
\caption{Numerical inputs used for the valence subtraction and pion-anchored mass units. The uncertainties are propagated numerically.}
\label{tab:scarab-inputs}
\begin{tabular}{lll}
\hline
Quantity & Value (MeV) & Role in the construction \\
\hline
\(m_u\) & \(2.16\pm0.07\) & current \(u\)-quark mass \\
\(m_d\) & \(4.70\pm0.07\) & current \(d\)-quark mass \\
\(m_s\) & \(92.90\pm0.70\) & current \(s\)-quark mass \\
\(M_{\pi^0}\) & \(134.9768\pm0.0005\) & neutral-pion anchor \\
\(M_{\pi^\pm}\) & \(139.57039\pm0.00018\) & charged-pion anchor \\
\hline
\end{tabular}
\end{table}

Each ordinary hadron is assigned independently. Multiplet membership, border color, connecting lines, and the desired shape of the final chart do not enter the minimization in Eq.~\eqref{eq:integer-search}. Mixed neutral mesons are treated component by component: the measured mass of the physical state is held fixed, while the valence subtraction and integer search are repeated for each displayed \(u\bar u\), \(d\bar d\), or \(s\bar s\) bookkeeping channel. The additional solid black outer border and the label ``M'' identify these tiles as flavor-component representations of a mixed state rather than as additional physical particles. There is one declared assignment constraint in the chart. The \(\Delta^{++}\) tile is placed at \(N_\pi=9\) with \((N_{uu},N_{dd},N_{\pm})=(0,0,9)\), so that all four \(\Delta\) charge states occupy the same decuplet row. Its model mass and residual are recalculated from that imposed triple, and the corresponding output-table entry is labelled as a manual override. No other hadron assignment is altered to enforce multiplet alignment. The revised detailed and classroom figures retain the literal \((I_3,N_\pi)\) row and column positions. Local offsets within crowded cells are used only to prevent overlapping tiles. Uncertainties are propagated by adding the input contributions in quadrature. Covariances among the derived pion units are neglected. For a selected triple, the model uncertainty includes the valence-quark input widths and the repeated pion-unit input widths. The residual uncertainty is obtained from the experimental and model uncertainties in the same manner. These propagated widths are included as diagnostics of the numerical construction. They do not turn the integer assignment into a precision determination of hadron masses.


\clearpage
\section{Simplified classroom chart}
\label{app:simplified-chart}

\begin{figure}[h!]
\centering
\IfFileExists{scarab_chart_classroom_circles.pdf}{%
\includegraphics[width=0.99\linewidth]{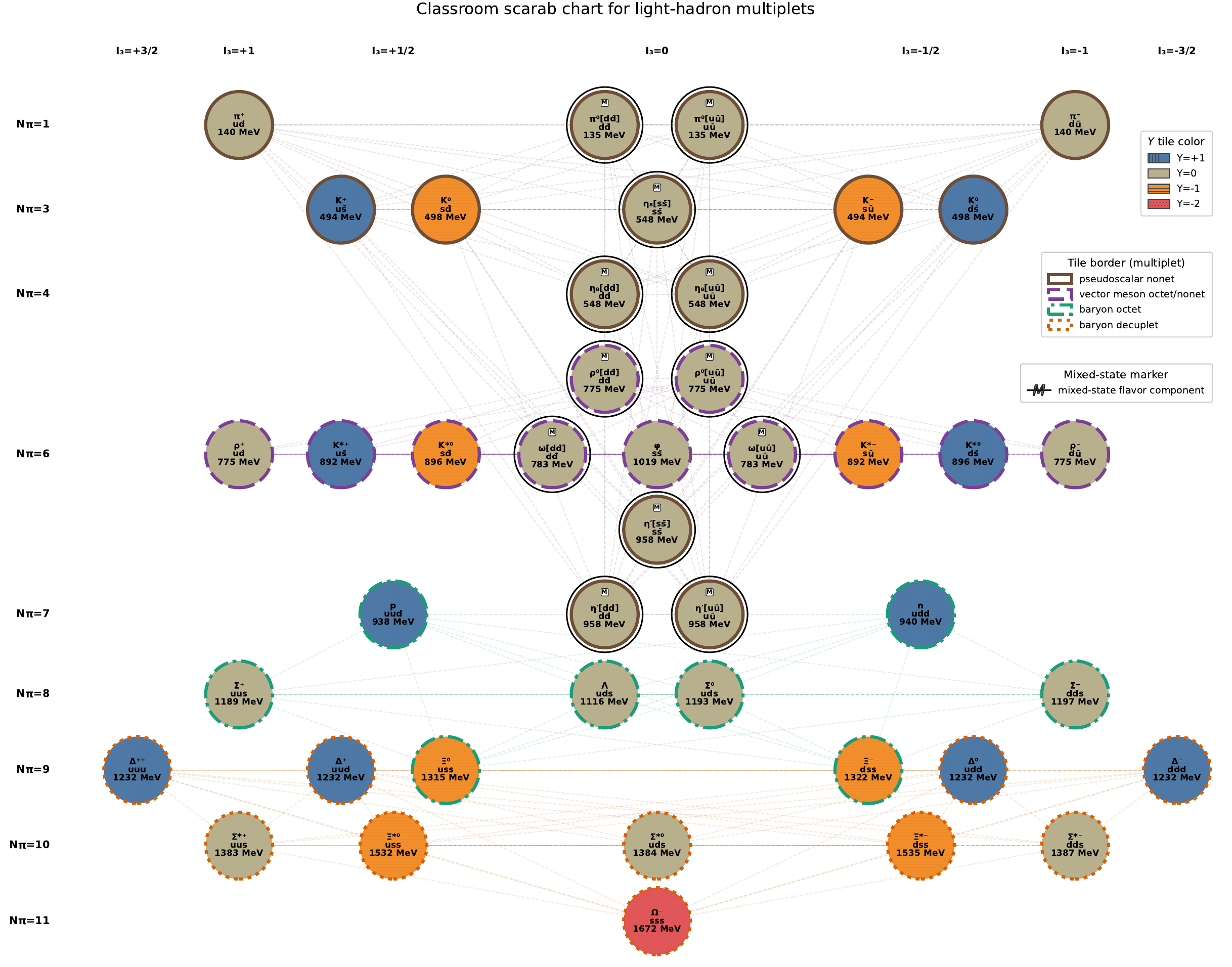}%
}{%
\IfFileExists{scarab_chart_classroom_circles.png}{%
\includegraphics[width=0.99\linewidth]{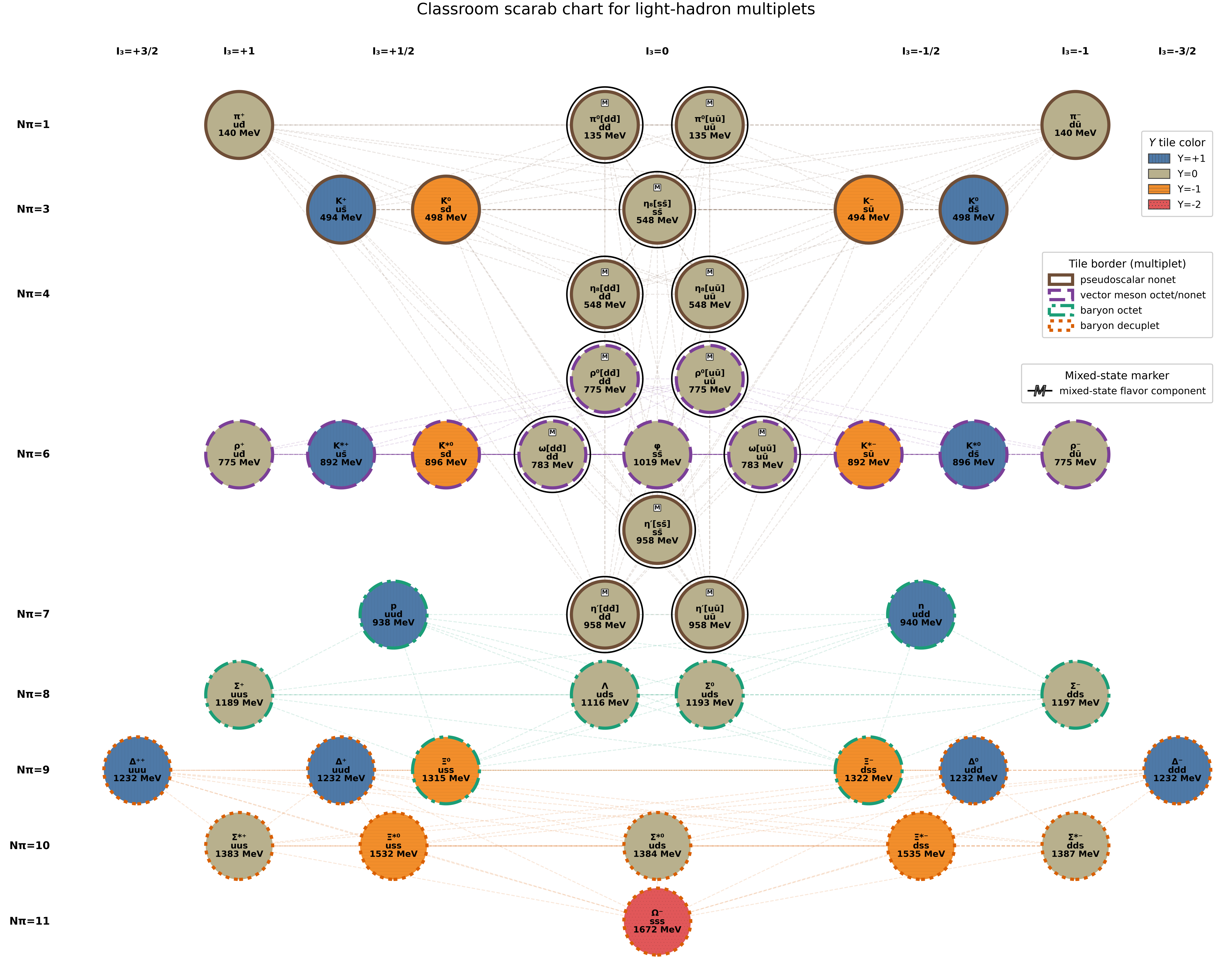}%
}{%
\fbox{\parbox{0.90\linewidth}{\centering
Placeholder for \texttt{scarab\_chart\_classroom\_circles.pdf}.\\
Place the simplified classroom chart in the manuscript directory or update
the filename in this appendix.}}%
}%
}
\caption{Simplified classroom scarab chart. Columns give \(I_3\), rows give
\(N_\pi\), tile fill colour and pattern give hypercharge \(Y\), and border
colour and line pattern give the multiplet class. The solid black outer ring and
the label ``M'' identify flavor-component representations of mixed neutral
mesons. Dashed connecting lines retain the multiplet relationships shown in
the complete chart.}
\label{fig:scarab-classroom}
\end{figure}

Figure~\ref{fig:scarab-classroom} presents the simplified version used in the
learning activity of Section~\ref{sec:classroom-activity}. It preserves the
scientific coordinates and redundant visual encodings of the complete chart,
but limits each circular tile to the particle or flavor-component symbol,
valence label, and measured mass. Tile fill colour and pattern encode
hypercharge, border colour and line pattern encode multiplet membership, and a
solid black outer ring with the label ``M'' identifies a displayed
flavor-component representation of a mixed neutral meson.

\end{document}